\documentclass[preprint,12pt]{elsarticle}

\usepackage{amssymb}
\usepackage{float}

\usepackage{multirow}
\usepackage{booktabs}
\usepackage[dvipsnames]{xcolor}
\usepackage{colortbl}
\definecolor{mypink}{rgb}{.99,.91,.95}
\definecolor{mygreen}{rgb}{.91,.99,.95}
\definecolor{myyellow}{rgb}{.98,.93,.80} 
\usepackage{pdfpages}
\journal{Journal of Nuclear Material}

\begin{document}

\begin{frontmatter}



\title{Effects of applied mechanical strain on vacancy clustering in FCC Ni}


\author{Shasha Huang}
	\author{Haohua Wen}
\author{Qing Guo}
\author{Biao Wang}
\author{Kan Lai$^{\ast}$}%
%\ead{laik3@mail.sysu.edu.cn}
\address{Sino-French Institute of Nuclear Engineering and Technology, Sun Yat-sen University, Zhuhai 519082, Guangdong, China}

\begin{abstract}
	Irradiation-induced vacancy evolution in face-centered cubic (FCC) Ni under mechanical strains was studied using molecular dynamics simulations. Applied hydrostatic strain $\varepsilon$ led to different stable forms of vacancy clusters, i.e., voids under $\varepsilon \ge +2\%$ and stacking fault tetrahedras (SFTs) under $\varepsilon \le 0$. Direct transitions between SFT and void revealed that increasing $\varepsilon$ magnitude facilitated the thermodynamic stability and dynamical evolution. The estimated free energy difference could well validate the dynamical simulations results by accounting for entropic contribution, which was revealed to play an important role in the thermodynamic stability of vacancy clusters in FCC Ni.
\end{abstract}


\end{frontmatter}

% \linenumbers

\section{Introduction}~\label{Introduction}
%结构材料很重要，对安全、经济性有影响
%结构材料遭受严酷环境：高温、高压、化学、辐射
In nuclear reactors, structural materials are usually exposed to an extremely severe environment, such as high pressure, high temperature, and high radiation doses, which lead to vacancies and interstitials inside the materials~\cite{comprehensive,fundamentals2016}. Compared with interstitials, vacancies can persist for a long time owing to a larger migration energy, and further form clusters~\cite{PhysRevB, KIRITANI200041, KIRITANI1991135}. Intensive studies including both experiments and numerical simulations have reported that voids and stacking fault tetrahedras (SFTs) are dominant vacancy-cluster forms in face-centered cubic (FCC) metals (e.g., Ni, Al, and Cu)~\cite{AIDHY201569, AIDHY2016137, OLSEN2016153, ZHAO201871, ZHANG201778, KADOYOSHI20073073}, which could be formed directly in the primary radiation damage~\cite{LI201660, Nordlund}. In addition, SFTs could also form as a result of dislocation loops gliding~\cite{WIRTH2000773, Silcox}, collapse of other vacancy clusters~\cite{Uberuaga2007,Xv2019,ZHAO201871}, and aggregation of vacancies~\cite{AIDHY2016137}. Both SFTs and voids could lead to adverse mechanical degradation, i.e., the presence of SFTs can cause hardening and embrittlement in metals whereas voids can induce swelling~\cite{ZHANG201727, SINGH2002159, CROSBY2014126, GHONIEM2000166, BRAILSFORD1972121}. Therefore, the formation and evolution behaviors of both void and SFT have attracted considerable attention. 

Several external factors are responsible for the vacancy clustering behaviors in FCC metals, e.g., temperature, vacancy concentration and irradiation technology~\cite{Xv2019, Uberuaga2007, Lounis2016, Mobility, KITAGAWA1985395, 1991PMagA}. Recent studies found that mechanical strains significantly affects the evolution of produced vacancies in FCC metals under irradiation, consequently affecting the clustering behaviors, e.g., stability~\cite{Xv2019, ZHANG201778}, concentration~\cite{FUJII2015281}, orientation~\cite{YE2016361, ye2017}, and structure~\cite{ZHAO201871, KADOYOSHI20073073}. However, because of the complex nature of the sources of mechanical strains exerted by structural materials, such as thermal stress, external loading, and strains induced by void swelling and solute precipitation~\cite{BRAILSFORD1972121, AUGER2000331}, the effects of applied mechanical strain on the stability and evolution behavior of void and SFT are still unclear. 
Moreover, the clustering of vacancies in FCC metals can exhibit very complex behavior.
A recent work on the stability of vacancy cluster in FCC Ni showed that even though taking into account the volumetric strains, an unexpected discrepancy was found between the energetic analysis using density function theory (DFT) at 0 K and thermodynamic stability using \emph{ab initio} molecular dynamics (AIMD) at finite temperature~\cite{ZHAO201871}. 
Another study in FCC Cu also showed that entropy change should be considered in the transformation process from void to SFT~\cite{Uberuaga2007}.

Therefore, in this study, by taking FCC Ni (a prototype system of a large group of Ni-based high entropy alloys~\cite{HEA}) as an example, we performed molecular dynamics (MD) simulation to study the formation process of vacancy clusters under the applied hydrostatic strains.
The stability of SFT and void under different strain conditions are analyzed from thermodynamic and energetic point of
view, which could help us understand microstructural evolution under irradiation in FCC metals.

\section{Method and simulation details}~\label{Methods}
\begin{figure}[ht!]
	\centering
	\includegraphics[width=1.05\linewidth]{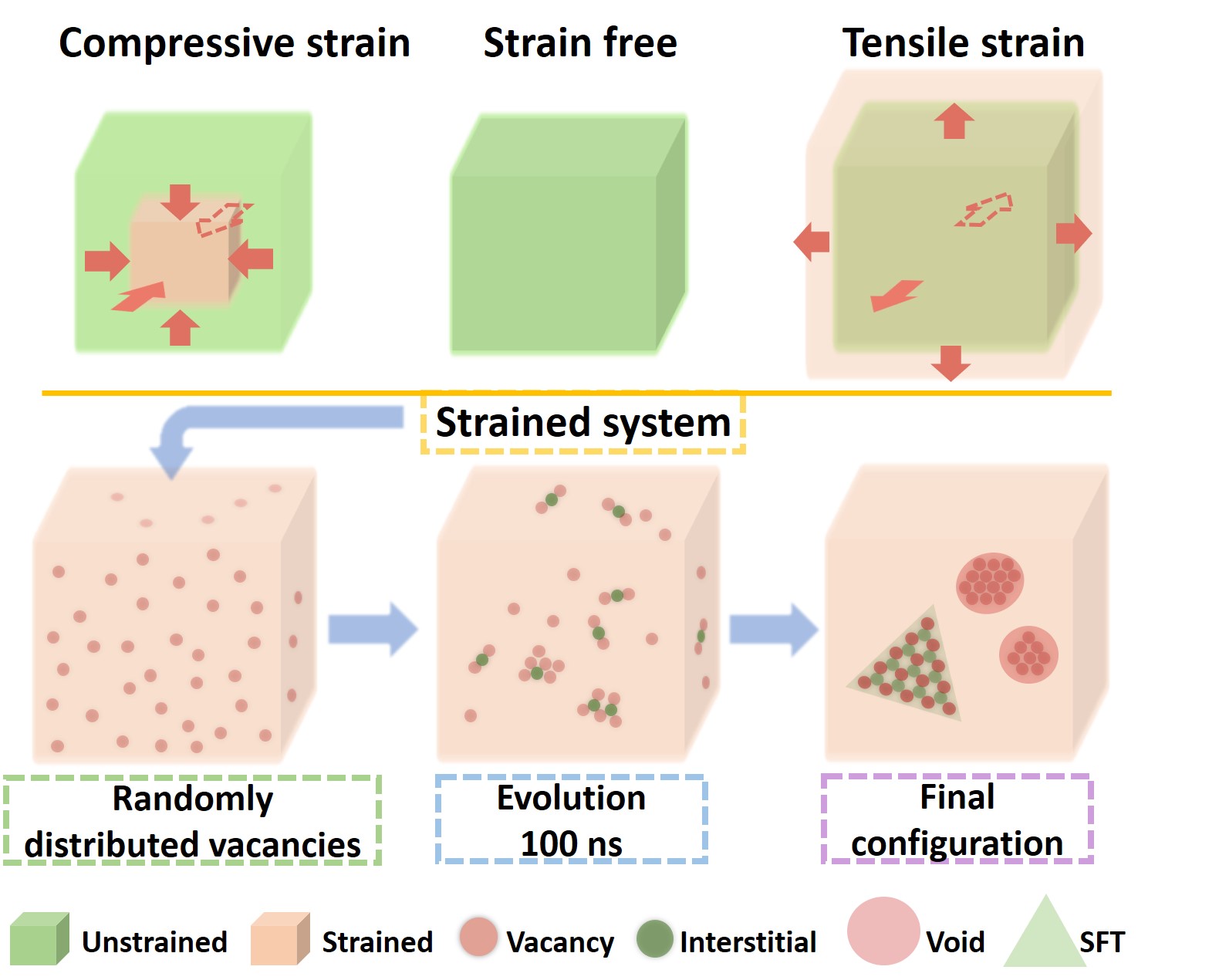}
	\caption{\label{Fig_1}
		Schematics of mechanical strains applied to FCC Ni crystal (top) and vacancy evolution (bottom). The light green box represents the original FCC crystal, the orange box the strained one, and the red arrow the action of the strains applied. Vacancies are represented by red spheres, whereas interstitials are represented by green spheres.
	}
\end{figure}
All results were obtained by MD simulations for FCC Ni using the Large-scale Atomic/Molecular Massively Parallel Simulator (LAMMPS) software package~\cite{lammps}. Bonny-2013 interatomic potential based on the embedded-atom method~\cite{EAMpotential} was adopted to describe the interatomic interaction, which has been confirmed to be suitable for the evolution of crystalline defects in FCC Ni~\cite{AIDHY201569,AIDHY2016137, Zhao2017}. The simulation cell contains 20$\times$20$\times$20 FCC unit-cells, with periodic boundary condition applied to avoid surface effect. Perfect FCC Ni crystal including 32,000 atoms was thermalized using isothermal-isobaric (NPT) ensemble at $1000$~K to obtain the stress-free equilibrium lattice constant, i.e., $a_0 = 3.6014$ \AA. 	
Next, the applied hydrostatic stains $\varepsilon$ was implemented by stretching or compressing the supercell along each dimension with the same percentage. Here, various $\varepsilon$ values were applied in the range of $-3 \% \le \varepsilon \le +3\%$, with $\varepsilon < 0$ corresponding to compression, and vice versa. For a specific simulation, after $200$ pico-seconds (ps) relaxation for the strained crystal at 1000 K using canonical (NVT) ensemble, certain numbers of vacancies were randomly introduced, to obtain the configuration of vacancy clustering in another $100$ nano-second (ns) MD runs. Here, the temperature and the initial number of vacancies were set to $1000$~K and $100$, respectively, in accordance with previous studies~\cite{AIDHY2016137}. Finally, defects were identified by displaced atom (D-A) analysis, which is widely adopted to identify SFT~\cite{VOSKOBOINIKOV2008385, article3}. 
As per the analysis, a sphere centered on an arbitrary perfect lattice site with a radius of 0.27 lattice constant ($1.0$~\AA)~\cite{AIDHY2016137} was used to define the spacing occupied by a regular atom; consequently, a sphere with a missing atom was identified as a vacancy, and any atom that did not locate within these spheres was identified as an interstitial.
Visual images of defect configuration were generated using the VMD software version 1.9.3~\cite{vmd}. In particular, at least $10$ independent MD simulations were carried out for each applied strain to check for statistical error.

\section{Results and discussion}\label{Results}
\begin{figure*}[ht!]
	\centering
	\includegraphics[width=0.9\linewidth]{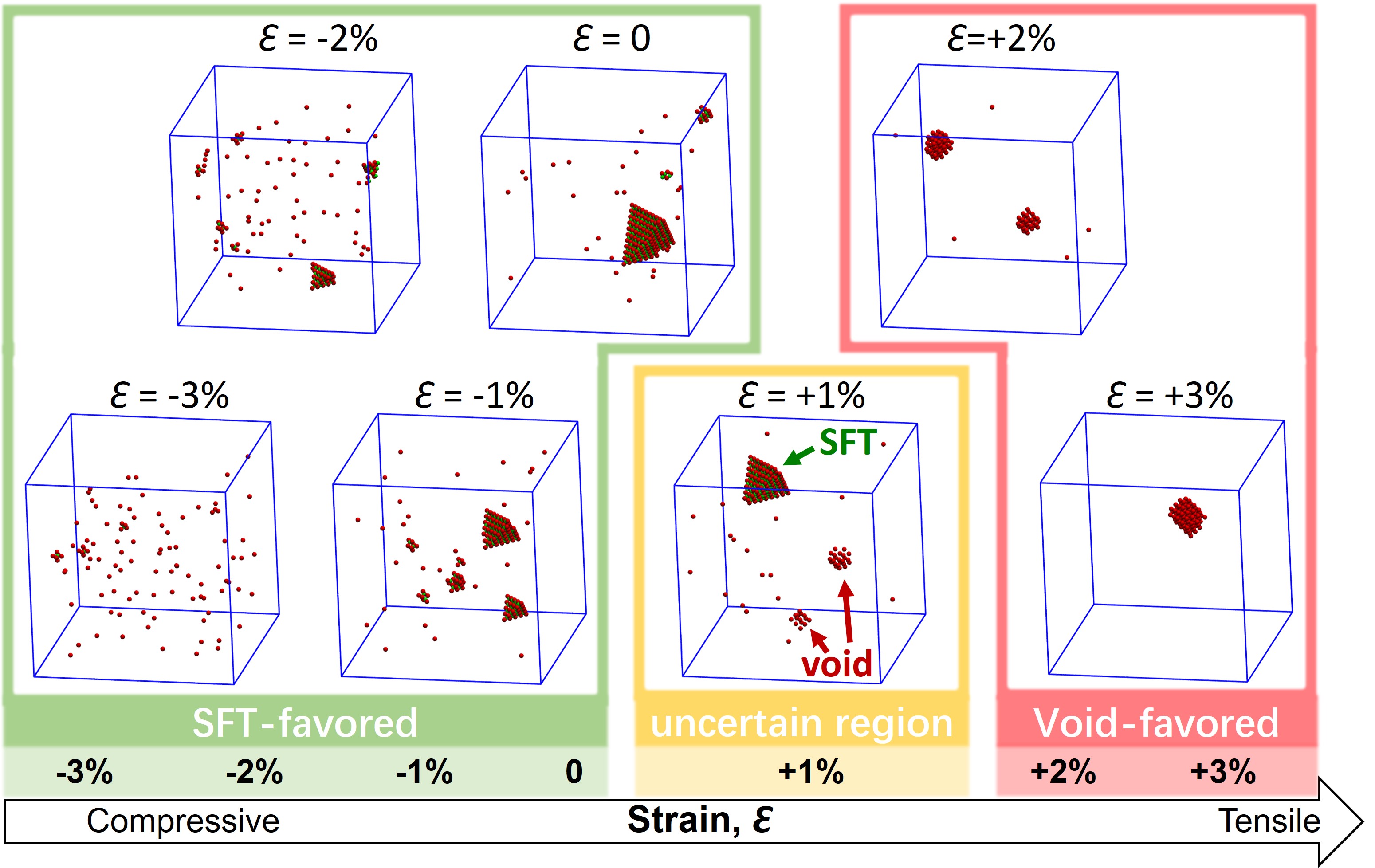}
	\caption{\label{Fig_2}
		Snapshots of vacancy clustering in FCC Ni after 100 ns MD runs under different hydrostatic strain $\varepsilon$ conditions. Vacancies are represented by red spheres, whereas interstitials are represented by green spheres.
	}
\end{figure*}
Fig.~\ref{Fig_2} plots the snapshot of vacancy clustering after 100 ns MD runs under various strains $\varepsilon$. In the case of $\varepsilon = 0$, SFTs were formed via vacancy-aggregation mechanism, and no voids are found, which is in good agreement with previous studies~\cite{AIDHY201569, AIDHY2016137}. Further, increasing compressive strain ($\varepsilon < 0$) led to a similar vacancy clustering behavior, i.e., no void, but smaller-sized SFTs. Such strain effect has been reported with FCC Cu~\cite{Xv2019}. In contrast, when sufficiently large tensile strains were applied, i.e., $\varepsilon \ge +2\%$, as shown in Fig.~\ref{Fig_2}, vacancies preferred to form voids, which enlarged along with tensile strains. In this regard, given a certain number of vacancy, either SFTs or voids would be formed, depending on the applied strain, e.g., for a system initially containing 100 vacancies as shown in Fig.~\ref{Fig_2}, $\varepsilon \le 0$ and $\varepsilon \ge +2\%$ respectively corresponded to the SFT and void-favored region, whereas $\varepsilon \sim +1\%$ was an uncertain region where SFT and void could exist independently or simultaneously (See detailed discussion in supplementary material Table S1).

\begin{figure}[htb!]
	\centering
	\includegraphics[width=1\linewidth]{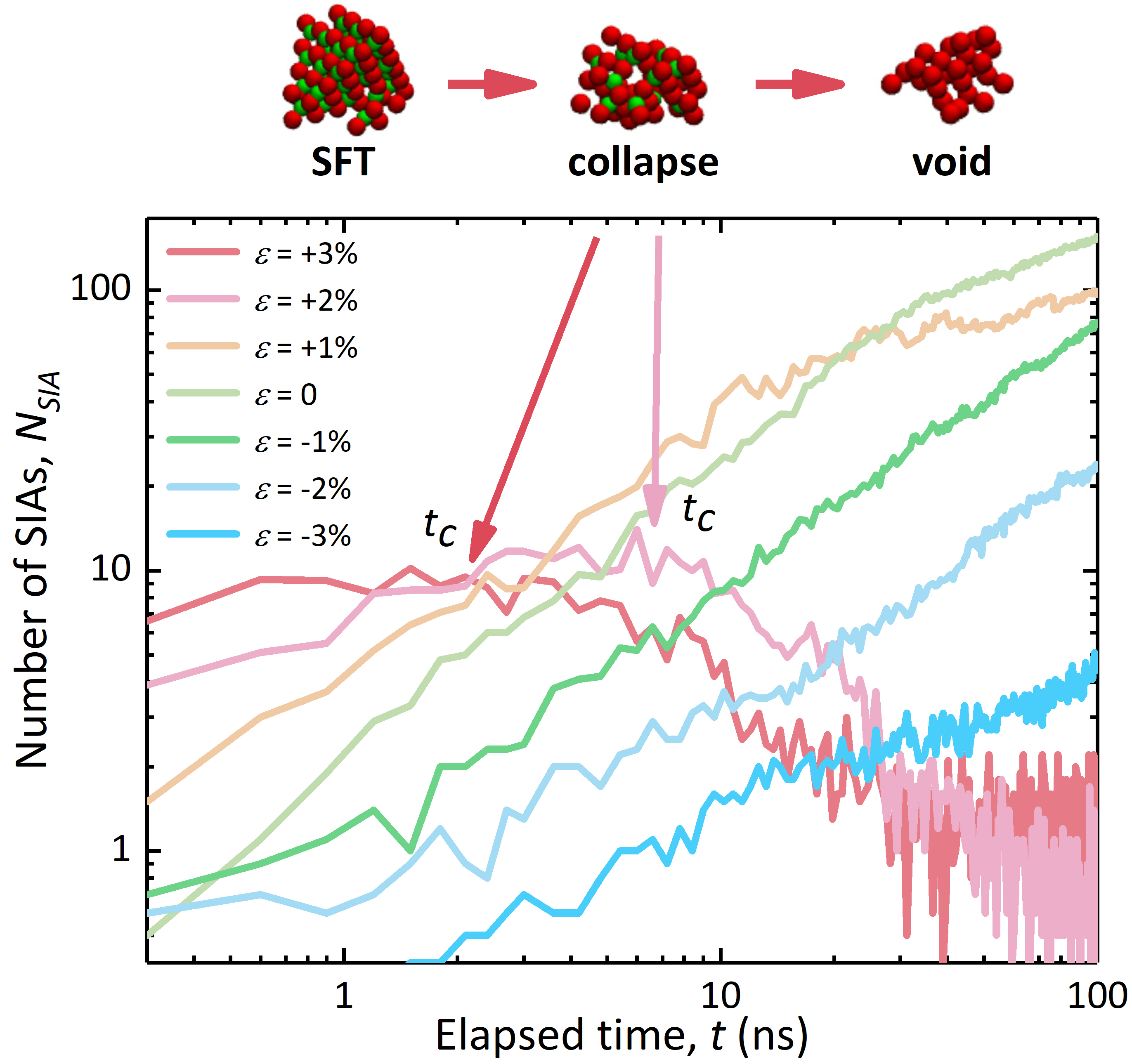}
	\caption{\label{Fig_3}
		Time-dependence of the number of SIAs ($N_{SIA}$) during vacancy clustering under various applied strains. Atomic view of SFT collapsing into void under tensile strain is presented on top (Red spheres represent vacancies and green spheres represent interstitials). 
	}
\end{figure}

Starting from the initial crystal configuration with 100 randomly distributed free-moving vacancies, stable defect configuration would be spontaneously formed, e.g., SFTs under $\varepsilon \le 0$ and voids under $\varepsilon \geq +2\%$ (See videos in supplementary files). It is noteworthy that during the formation of SFT by the diffusion-based vacancy-aggregation mechanism~\cite{AIDHY2016137}, the aggregated vacancies prompt nearby lattice atoms to move towards clusters to self-form a Frenkel pair~\cite{Uberuaga2007}. The number of self-interstitial atoms (SIAs) ($N_{SIA}$) defined using the D-A approach~\cite{VOSKOBOINIKOV2008385, article3} was revealed to increase with the growth of SFT. In addition, there is a vacuum space inside the void structure,which contained no SIAs. Therefore, the time-evolution of $N_{SIA}$ could be used to characterize the formation process of SFT. As shown in Fig.~\ref{Fig_3}, in the SFT-favored conditions, i.e., $\varepsilon \le 0$, $N_{SIA}$ increased as time $t$ elapsed, indicating that vacancies aggregated and formed SFTs. A different process was revealed in the void-favored conditions, i.e., $\varepsilon \ge +2\%$, where two successive evolution stages were found. For the early stage, i.e., $t < t_c$ ($t_c$ denoting the instant when $N_{SIA}$ significantly started to decrease), similar increasing behavior of $N_{SIA}$ was observed, indicating that small SFT-like structures were probably formed. Subsequently, these SFTs collapsed into void gradually, as the SIAs inside recombined with the neighboring vacancies to annihilate one another, thus leading to the reduction of $N_{SIA}$ at the second stage, i.e., $t > t_c$. Further, the critical time $t_c$ was found be sensitive to the applied strain, e.g., $t_c=8.4\pm1.0$ ns at $\varepsilon = +2\%$ and $t_c=3.4\pm0.5$ ns at $\varepsilon = +3\%$. In addition, as vacancy clustering is a diffusion-controlled process, the applied tensile (compressive) strains could facilitate (suppress) vacancy diffusion, thus causing faster (slower) growth of SFTs or voids. In this regard, the applied mechanical strains led to the different stable configurations and the dynamical evolution behavior of vacancy clustering (see also the mean square displacement provided in supplementary material Fig. S1).

%\section{Discussion}~\label{Discussion}

The abovementioned forms of vacancy clusters at finite temperatures were speculated to be a meta-stable state, and a high energy barrier between SFTs and voids states under certain applied strains might exist; thus, requiring a sufficient long evolution process for the pre-formed structure to directly transition into another structure, which exceeds the limit of current MD simulations~\cite{Uberuaga2007}. To check this conjecture, direct transitions from SFT to void and from void to SFT were simulated as follows.

To obtain the initial state of a direct transition simulation, a specific vacancy cluster (either SFT or void) was artificially constructed inside a strained crystal system following the approach detailed described in Ref.~\cite{ZHAO201871,Xv2019}. Next, energy minimization at $0$~K was performed to optimize such defect structure. The strained system was subsequently relaxed at 1000 K using canonical ensemble for at least 2 ns to simulate the direct transition processes. In addition, as the size of vacancy cluster, which is equivalent to the number of vacancies involved, is a predominant factor for thermodynamic stability~\cite{Uberuaga2007,ZHAO201871}, various SFT and void sizes were considered to comprise magic or non-magic numbers of vacancies (i.e., $20$, $35$, $45$, and $55$). Further, $15$ independent simulations for each case were conducted to limit statistical error.

\begin{table*}[tb!]
	\centering
	\renewcommand\arraystretch{1.2}
	\caption{Transition time (ps) for the process between SFT and void under different strain conditions, each calculated from an average of $15$ individual simulations. If transition did not occur, the data are denoted by ``--''. In some cases, where the transformation only happened in a few of the $15$ simulations, the data are denoted by ``PC'', meaning partly changed.} 

\resizebox{\textwidth}{!}{
	\begin{tabular}{ccccccccc}
		\toprule
		\multirow{2}{*}{$\varepsilon$}&\multicolumn{2}{c}{20 vacancies}&\multicolumn{2}{c}{35 vacancies}&\multicolumn{2}{c}{40 vacancies} &\multicolumn{2}{c}{55 vacancies}\\
		\cline{2-9}
		&SFT to void&void to SFT&SFT to void&void to SFT&SFT to void&void to SFT&SFT to void&void to SFT\\
		\midrule
		\rowcolor{mypink}
		+3\%&$0.9\pm0.1$& -- &$6.3\pm0.5$& -- &$10.1\pm1.6$& -- & $20.6\pm2.0$&-- \\
		\rowcolor{mypink}
		+2\%&$7.5\pm3.3$& -- &$36.7\pm3.4$&-- &$41.2\pm2.5$&--& $457\pm118$& --\\
		\rowcolor{myyellow}
		+1\%&PC& PC & -- &-- &PC& --& --& --  \\
		\rowcolor{mygreen}
		0& --  & $21.6\pm2.1$  & -- & $119\pm41$ & --& PC  & -- & --  \\
		\rowcolor{mygreen}
		-1\%&--  & $12.1\pm0.8$  & --& $12.1\pm0.7$ & -- & $20.3\pm1.2$  &-- & $71.7\pm10.8$  \\
		\rowcolor{mygreen}
		-2\%& --  & $2.6\pm0.5$  & --& $7.8\pm0.5$ & --& $14.2\pm0.7$  & -- & $27.9\pm5.7$  \\
		\rowcolor{mygreen}
		-3\%&--  & $1.6\pm0.2$  & -- & $4.0\pm0.3$ & -- & $7.8\pm0.3$ & -- & $11.8\pm0.7$  \\
		\bottomrule
		\label{Formation} 
	\end{tabular}}
\end{table*}

As listed in Table~\ref{Formation}, transition from SFT to void was not observed at the SFT-favored strain region ($\varepsilon \le 0 $) as illustrated in Fig.~\ref{Fig_2}. Similarly, transition from void to SFT was observed at the void-favored strain region ($\varepsilon \ge +2\% $). This is consistent with the observation in Fig.~\ref{Fig_2} that applied strain led to different stable vacancy clusters at finite temperatures. Otherwise, if an ``unstable'' vacancy cluster was initially constructed at its ``un-favored'' applied strain, it would spontaneously transfer into the ``stable'' form; therefore, the corresponding transition occurred in a specific time-interval (the so-called \emph{transition time}), ranging from several to hundreds of ps, as shown in Table~\ref{Formation}. In such conditions, larger magnitude of the applied strain and smaller cluster size led to shorter transition time. However, exceptions (in which voids did not transform into SFT within 2 ns) were found under $\varepsilon=0$ with large voids containing $40$ and $55$ vacancies, indicating the presence of a high energy barrier. In addition, at an uncertain region of $\varepsilon \sim +1\%$, structural transition between SFT and void appeared to be rare, the detail of this mechanism requires further investigation. According to the classical nucleation theory, high energy barrier decreases the transition rate, thereby resulting in longer transition time. In this regard, the applied strains suppressed the energy barrier between SFTs and voids, thus facilitating the formation and thermodynamic stability of vacancy clustering.  

\begin{figure}[htb!]
	\centering
	\includegraphics[width=1.05\linewidth]{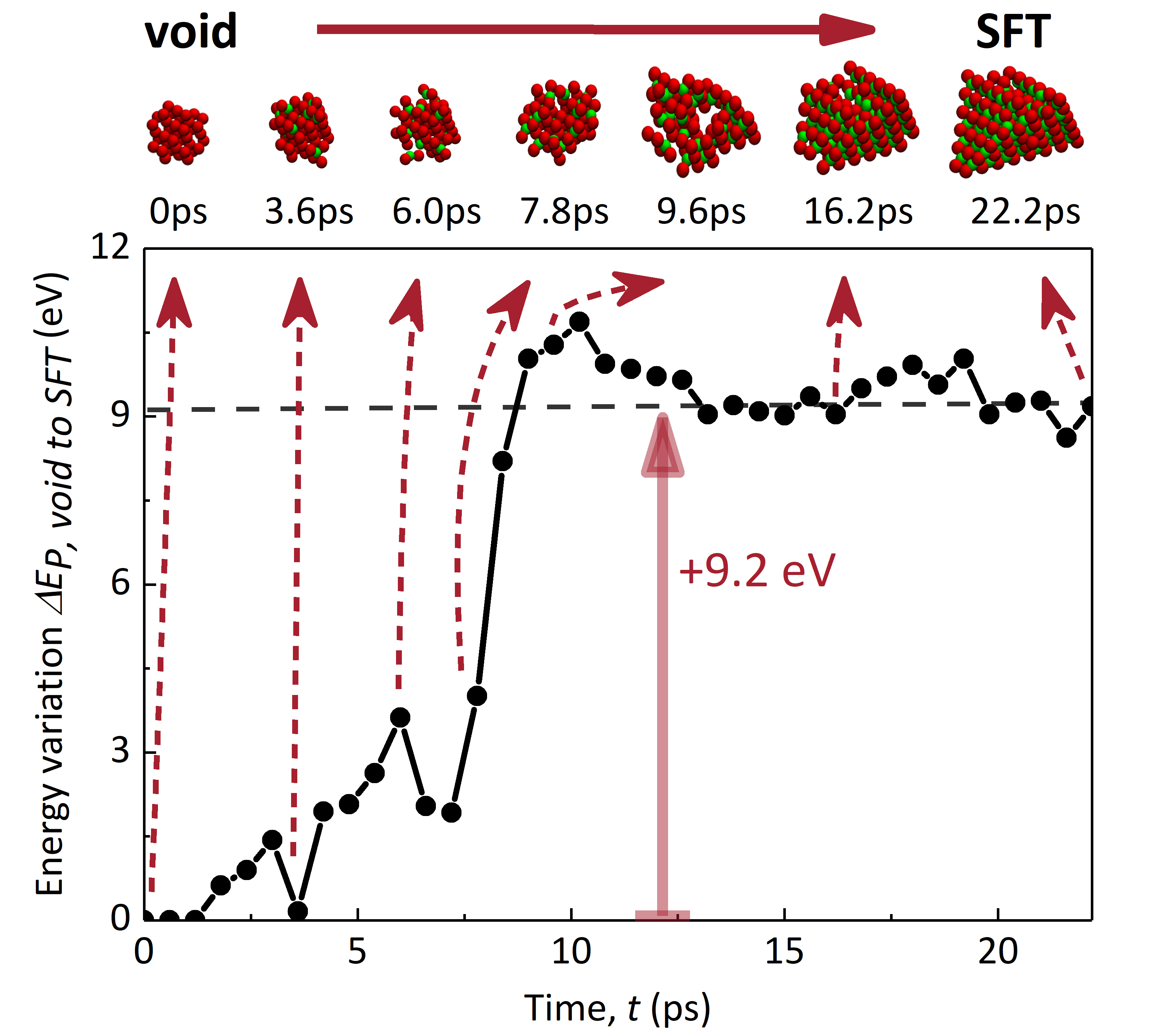}
	\caption{Time-dependence of potential energy variation during the transition process form void to SFT in the strained system initially containing $40$ vacancies under $-1\%$ strain. Atomic view of the vacancy clustering structure is presented on top with the time elapsed (vacancies are represented by red spheres, whereas interstitials are represented by green spheres).}\label{Fig_4}
\end{figure}

From the energetic point of view, spontaneous vacancy clustering in metals occurs owing to the large positive binding energies arising from the elastic interactions between defects. Hence, the clustering form with lower energy is more stable, and in fact becomes a criterion widely adopted in numerical modeling via the DFT and MD methods. However, recent \emph{ab initio} results based on the DFT~\cite{ZHAO201871} suggested that the formation of void should be energetically favorable at $0$~K in FCC Ni, even in the case of applied strains, which is inconsistent with their \emph{ab initio} molecular dynamics simulations and our simulation results at finite temperatures. Taking a typical direct transition from void to SFT at the SFT-favored strain region (e.g., $\varepsilon = -1\%$) as an example, the time-dependence of potential energy variation in the strained system containing $40$ vacancies during the 
transition process was traced to confirm the validity of such criterion. As plotted in Fig.~\ref{Fig_4}, potential energy continuously increases with the transition of void into SFT, where an energy increment of $\sim+9.2$~eV,	against expectation, occurred at the end of this transition. This result suggested that such criterion was insufficient to elucidate the stability of vacancy clusters at finite temperatures (more information on energy difference with cluster containing 3 to 100 vacancies is provided in supplementary material Fig. S2). Indeed, the discrepancies between prediction using energy criterion and MD simulation results have been reported, and they are probably caused by the temperature effect (lattice vibration)~\cite{ZHAO201871} or entropic contributions~\cite{Uberuaga2007}.

In the following analysis, entropic contribution was considered to examine the stability of vacancy clusters under applied strains. Accordingly, entropy changes with volumetric change in an isothermal system could be estimated approximately by $\Delta S=\alpha B \Delta V $, where $\alpha$ is the volumetric expansion coefficient, $B$ is the bulk modulus~\cite{Uberuaga2007,entropy2}, and $\Delta V$ is the difference in volume filled by Ni atoms between void and SFT states. Note that the effective volume occupied by the Ni atoms inside FCC crystal with SFTs is larger than that with voids~\cite{Uberuaga2007}, and the resulting entropy increment will enhance the thermodynamic stability of SFTs compared with that of voids. In this regard, in NVT ensemble, the Helmholtz-free energy difference ($\Delta F$) between states of SFT and void in a strained system could be calculated as follows: $\Delta F=\Delta U-T\Delta S$. In the current study, we adopted $\alpha=6.36\times10^{-5}$~/K and $B=1.73\times10^{11}$~Pa, and assumed $\Delta V = 0.8 \times N_{v} \times \Omega$, where $\Omega$ is the atomic volume ($\Omega=(1+\varepsilon) \times 11.7~\AA^{3}$) and $N_v$ is the number of vacancies involved. The energy difference $\Delta U$ was estimated as the difference in the lowest potential energy $\Delta E_{P}$ of SFT and void in several competitive structures with energy minimization ~\cite{ZHAO201871}, where kinetic energy is assumed to keep constant between two states under canonical ensemble. In the vacancy cluster with $N_v=100$ under strain-free condition at $1000$~K, compared with void, the formation of SFT led to an entropic contribution of $-T\Delta S \sim -60.5$~eV, which is sufficiently large to eliminate the corresponding energetic increment, i.e., $\Delta U \sim +43.6$~eV, thus significantly decreasing the free energy, i.e., $\Delta F_{SFT-void} \sim -16.9$~eV. That is why SFT was more stable in MD simulations at finite temperatures under $\varepsilon = -1\%$.    
\begin{figure}[tb]
	\centering
	\includegraphics[width=1.03\linewidth]{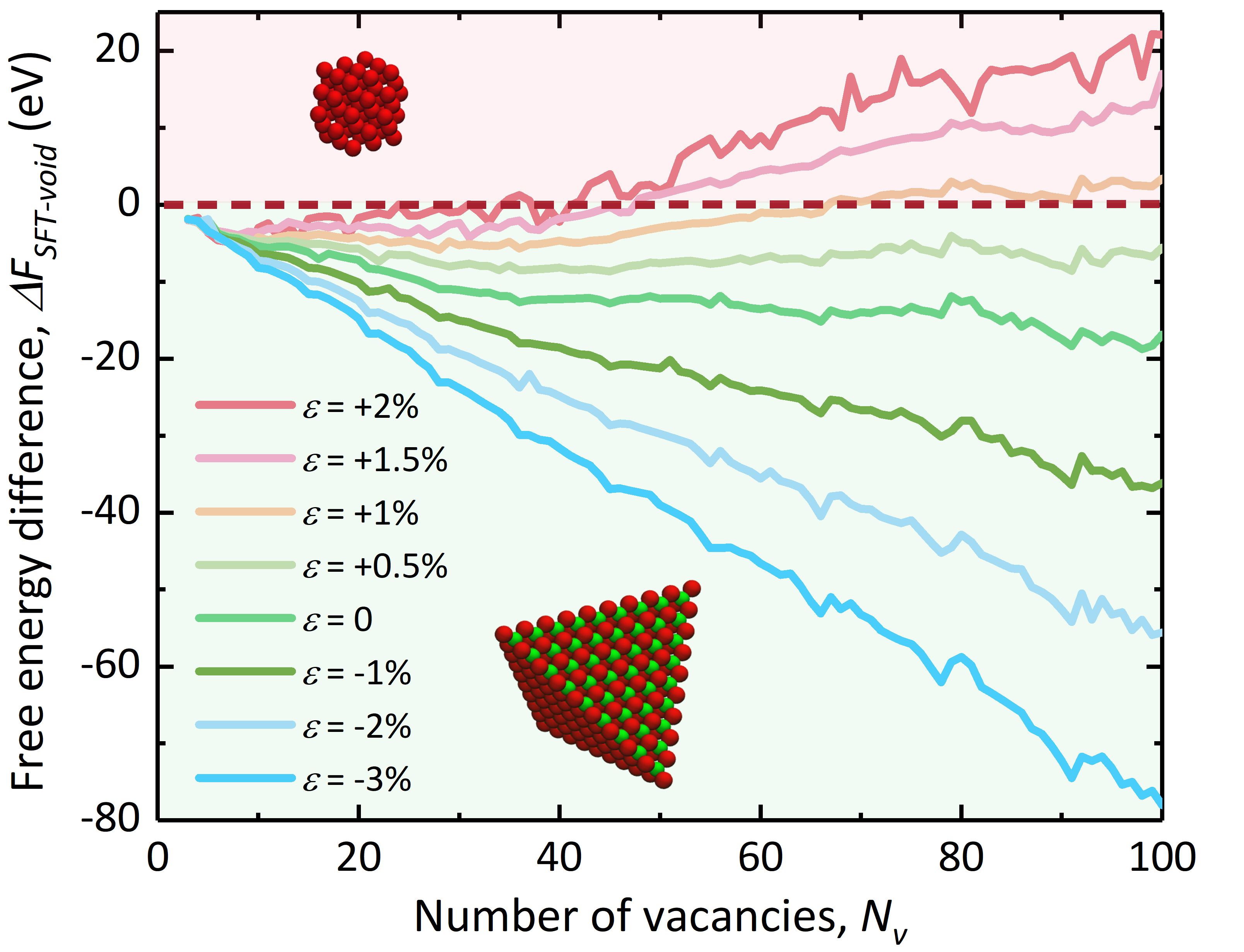}
	\caption{ Free energy difference of voids relative to SFTs as a function of vacancy cluster size in various strains applied.}
	\label{Fig_5}
\end{figure} 
\begin{figure}[tb]
	\centering
	\includegraphics[width=1.0\linewidth]{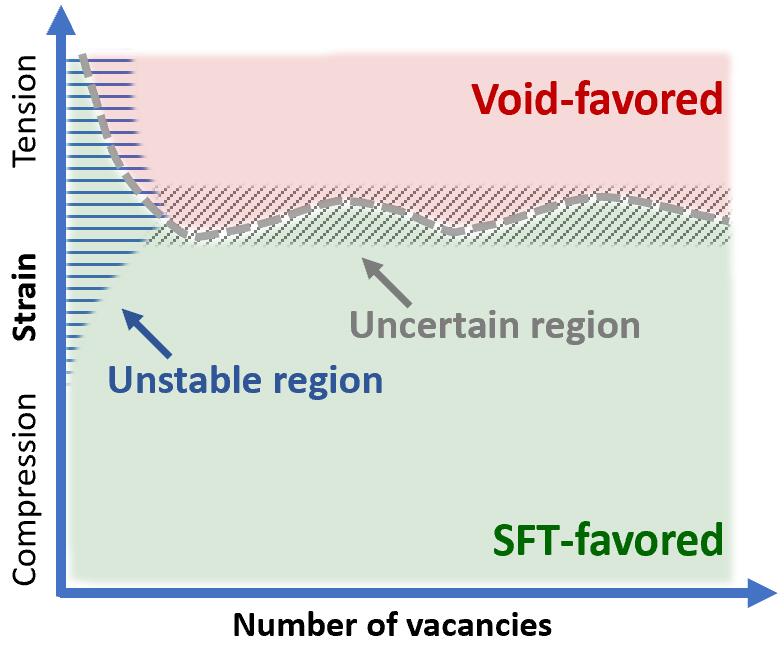}
	\caption{ Schematic illustration of the effect of strain on the stability of vacancy clusters. }
	\label{Fig_6}
\end{figure}    
Consequently, the free energy difference in the strained system between SFT and void, $\Delta F_{SFT-void}$ (simplified as $\Delta F$ below), was calculated as a function of $N_v$. The results are presented in Fig.~\ref{Fig_5}, where $\Delta F < 0$ indicates that SFT is more stable, and vice versa. Although the estimation of $\Delta F$ was not rigorous, the results shown in Fig.~\ref{Fig_5} are roughly consistent with the MD simulation results in the foregoing: under strain-free condition and compressive strain (e.g., $\varepsilon \le 0$), $\Delta F < 0$ and thus SFTs were stable; in contrast, under large tensile strain (e.g., $\varepsilon\ge+2\%$), $\Delta F > 0$ and voids were present. Schematics in Fig.~\ref{Fig_6} summarize the effects of applied strains on the stability of vacancy clustering with different sizes. In addition, as the magic numbers of the perfect SFTs and voids are different, the curve in Fig.~\ref{Fig_5} is somewhat serrated. Thus, under low tensile strain where $\Delta F$ was very small (e.g., $\varepsilon=+1\%$, brown line in Fig.~\ref{Fig_5}), the curve fluctuated around zero, indicating that the relative stability of SFT and void was size-dependent under $\varepsilon=+1\%$ (namely ``uncertain region'' in Fig.~\ref{Fig_6}). Further, under tensile strain, where void was generally considered more stable, there was a small size region where SFT was more stable (e.g., $\varepsilon=+3\%$ and $N_{v}<30$). During the formation process of large-sized void, small-sized SFTs would be formed in the early stage, grow by capturing more nearby vacancies, and then start to collapse into void when it reached a critical size (see also in Fig.~\ref{Fig_3}). Therefore, small-sized SFTs could possibly act as a nucleation center at the beginning of the void formation process. Moreover, there was also a region (namely ``unstable region'' in Fig.~\ref{Fig_6}) where both small SFTs and voids were not stable owing to the low binding energy (less than $10$~eV, as shown in supplementary material Fig. S3) and could be dissolved by thermal activation at 1000 K. Accordingly, we assessed the stability of small-sized SFTs and voids ($N_{v}<8$) in MD simulations and found that all the clusters were dissolved within $1$~ns.

\section{Conclusion}~\label{Conclusions}
In conclusion, we performed MD simulations to study the effects of applied mechanical strains on the stability and evolution of vacancies clustering in FCC Ni. It was observed that applied mechanical strains led to different stable configurations, i.e., voids were more stable under tensile strain, whereas SFTs were more favorable under compressive strain. 
We demonstrate the formation of SFT and void via different vacancy agglomeration mechanisms under different strain conditions. 
The stability of vacancy clusters was enhanced by large magnitude of applied strains. 
Moreover, the thermodynamic stability of these clusters is
also crosscheck by simulating direct transformation.
Further, our energetic analysis suggested that entropic contribution plays a crucial role in the stability of vacancy clusters in FCC Ni at finite temperature. 
It is known that the mechanical strain in irradiated materials which could affect vacancy clustering is dynamic and complex. 
On the other hand, the formation of vacancy clusters can lead to swelling (i.e., in the form of void) or hardening (i.e., in the form of SFT), thus could be sources of mechanical strain.
Therefore, it is important that these clustering behaviors be introduced in higher-level models developed for the long-term evolution of irradiation damage accumulation.

%%%%%%%%%%%%%%%%%%%%%%%%%%%%%%%%%%%%%%%%%%%%%%%%%%%%%%%%%%%%%%%%%%%%%%%%%%%%%%%%%%%%%%%%%%%%%%%%%%%%%%%%%%%%%%%%%%%%%%%%%%%%%%%
\section*{Acknowledgements}
\label{Acknowledgements}
The authors gratefully acknowledge Yingjie Tang, Guo Zhang, Fan Ye, Anruo Zhong, Jiewei Wu, and Zhibo Zhang for insightful discussions. This work was supported by the National Natural Science Foundation of China under Grant No. 11602311, 
%感谢国家青年基金
the Natural Science Foundation of Guangdong Province, China, 
under Grant No. 2014A030310165, 
%感谢省博士启动基金
the Fundamental Research Funds for the Central Universities under Grant No. 45000-31610018,
%感谢高校基本科研业务费 45000-31610018
Guangzhou Science and Technology Project No. 201707020002, 
%感谢某广州项目
and 
the Special Program for Applied Research on Super Computation of the NSFC-Guangdong Joint Fund (the second phase). 
%感谢天河二号



%% References
%% Usage of ~\cite is as follows:
%%   ~\cite{key}         ==>>  [#]
%%   ~\cite[chap. 2]{key} ==>> [#, chap. 2]
%%


\bibliographystyle{elsarticle-num-names}
%\bibliography{Ref_Ni}
\bibliography{Ref_Ni_new}






% \end{thebibliography}

%\clearpage



%% References







% \end{thebibliography}
	

\section*{Supplementary material}

# Supplementary material

# Effects of applied mechanical strain on vacancy clustering in FCC Ni

Shasha Huang, Haohua Wen, Qing Guo, Biao Wang, Kan Lai*

Sino-French Institute of Nuclear Engineering and Technology, Sun Yat-Sen University, Zhuhai 519082, Guangdong, China.
E-mail: laik3@mail.sysu.edu.cn (Kan Lai*)

For the uncertain region around +1%, we performed additional simulations under +0.75, +1, and +1.25% tensile strain, with 20 independent MD simulations for each. The final configurations (shown in Table S1) suggested that the uncertain region where SFT and void could exist independently or simultaneously was quite narrow.

Table S1 Proportion of different types of vacancy cluster in Ni under different strains.

| Strain value $\varepsilon$ | +0.75% | +1% | +1.25% |
| --- | --- | --- | --- |
| SFT : coexist : void | 20 : 0 : 0 | 6 : 2 : 12 | 0 : 0 : 20 |

During the vacancy evolution, at the early stage ($t < 20$ns), the isolated vacancies rapidly bound with the nearby vacancies formed small clusters, which diffused and grew at a fast rate. At the later period ($t > 20$ns), the gradually formed large clusters moved and grew at a slower rate (as shown by mean square displacement provided in Fig. S1). This phenomenon was consistent with the previous finding that diffusion coefficient is lower for larger clusters [1]. By incorporating the effect of the strain, it was revealed that the diffusion of vacancies increased as system volume increased. Therefore, the clustering of vacancies was facilitated under tensile strain, but suppressed under compressive strain, thereby resulting in the difference in cluster size observed at the end of simulations (see in Fig.1).

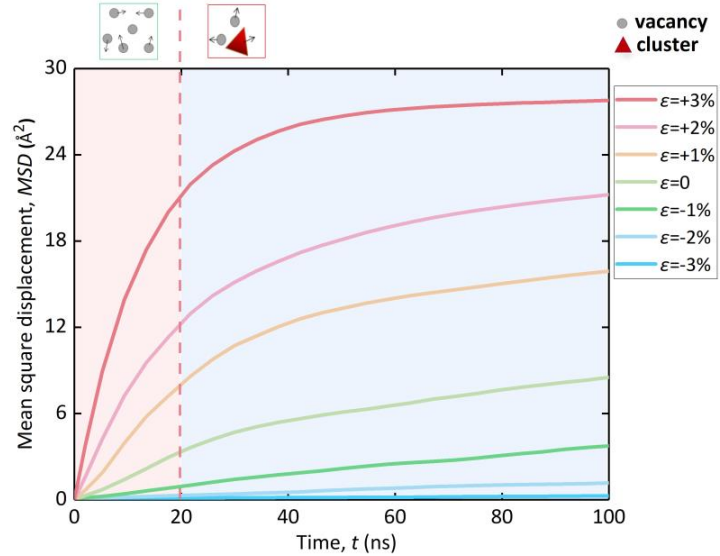

**Fig. S1**. Different stages of evolution illustrated by mean square displacement of atoms under different strain as a function of time.

Our MD simulations provided inconsistent results for the potential energy difference between SFT and void containing 3 to 100 vacancies (shown in Fig. S2), suggesting that the energetic analysis based only on potential energy was insufficient for understanding the stability of vacancy clusters at finite temperature.

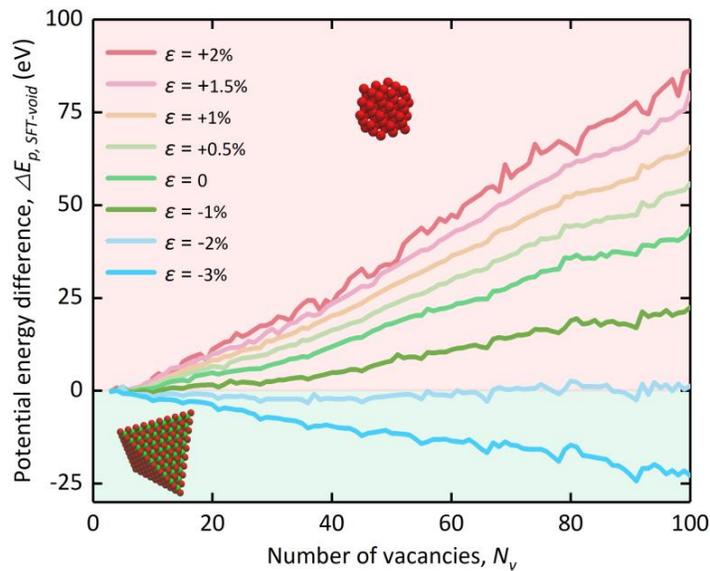

**Fig. S2**. Potential energy difference between SFT and void under different strain as a function of vacancy cluster size.

The binding energies of void and SFT increased as the vacancy cluster size increased. However, small clusters ($N < 8$) with low binding energies ($E_b < 10$ eV) can easily dissolved via thermal activation at 1000 K.

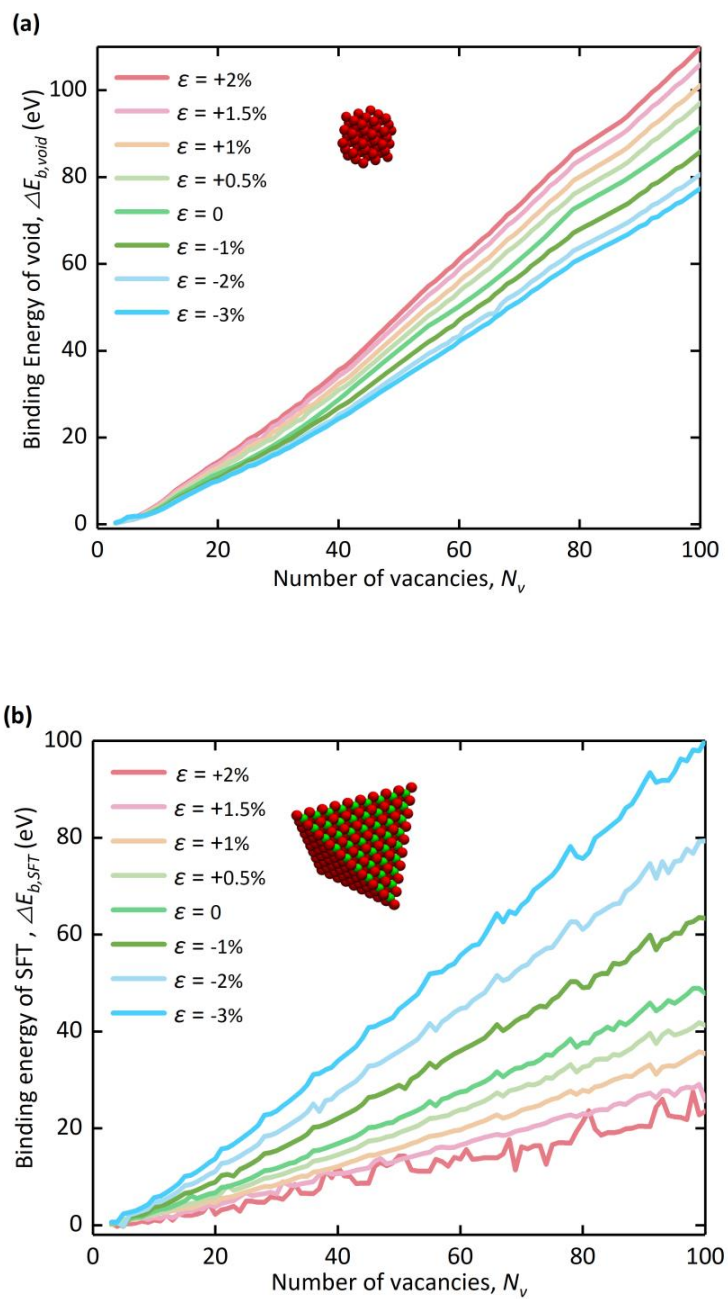

**Fig. S3**. Binding energy of void (a) and SFT (b) under different strain as a function of vacancy cluster size.



\end{document}